\documentclass[prl,nofootinbib,twocolumn]{revtex4}
\usepackage{amsmath,amssymb,bm,graphicx}
\usepackage[utf8]{inputenc}
\usepackage[T1]{fontenc}
\usepackage{mathptmx}

\newtheorem{Theorem}{Theorem}

\newcommand{\beq}{\begin{equation}}
\newcommand{\eeq}{\end{equation}}
\newcommand{\bear}{\begin{eqnarray}}
\newcommand{\ear}{\end{eqnarray}}
\newcommand{\bdm}{\begin{displaymath}}
\newcommand{\edm}{\end{displaymath}}

\def\oper{{\mathchoice{\rm 1\mskip-4mu l}{\rm 1\mskip-4mu l}
{\rm 1\mskip-4.5mu l}{\rm 1\mskip-5mu l}}}
\def\<{\langle}
\def\>{\rangle}

\begin{document}

%\title{\textbf{Non-Markovian quantum evolution from the dynamics of classical-quantum states}}
\title{\textbf{On the hybrid Davies like generator for quantum dissipation}}%: classical vs. quantum}}
\author{Dariusz Chru\'sci\'nski}
\affiliation{ Institute of Physics, Faculty of Physics, Astronomy and Informatics \\  Nicolaus Copernicus University,
Grudzi\c{a}dzka 5/7, 87--100 Toru\'n, Poland\\ email: darch@fizyka.umk.pl}

%\affiliation{Institute of Physics, Faculty of Physics, Astronomy and Informatics, \\ Nicolaus Copernicus University,
%Grudzi\c{a}dzka 5/7, 87--100 Toru\'n, Poland\\
%}

%\pacs{03.65.Yz, 03.65.Ta, 42.50.Lc}

\begin{abstract}

We provide a class of quantum evolution beyond Markovian semigroup. This class is governed by a hybrid Davies like generator such that dissipation is controlled by a suitable memory kernel and decoherence by standard GKLS generator. These two processes commute and both of them commute with the unitary evolution controlled by the systems Hamiltonian. The corresponding memory kernel gives rise to semi-Markov evolution of the diagonal elements of the density matrix. However, the corresponding evolution needs not be completely positive.  The role of decoherence generator is to restore complete positivity. Hence, to pose the dynamical problem one needs two processes generated by classical semi-Markov memory kernel and purely quantum decoherence generator. This scheme is illustrated for a qubit evolution.

\end{abstract}

\maketitle

{\bf A simple generalization of quantum Markovian semigroup is proposed. The generator of Markovian semigroup in the weak coupling limit  has the characteristic form derived by Davies (so called Davies generator) and consists of three commuting parts: Hamiltonian part responsible for unitary evolution, dissipative part and pure decoherence part. To include possible memory effect due to he nontrivial system-environment interaction we propose to replace Markovian dissipation by non-Markovian generator represented by non-local memory kernel. In general the corresponding evolution is no longer physically admissible since it is represented by a map which is not completely positive and trace preserving. In our scheme the memory kernel gives rise to purely classical dissipative process (so called semi-Markov process) for diagonal elements of the density operator interpreted as populations of the corresponding energy levels. Hence, whenever one restricts to diagonal elements the process  is physically legitimate. However, off-diagonal elements (quantum coherences) may cause the entire evolution not legitimate. Interestingly, it turns out that the additional Markovian generator responsible for pure decoherence may wash out all unphysical terms making the evolution perfectly physically admissible. This way the total process is described  by a hybrid generator, that is, an appropriate pair of semi-Markovian dissipation and Markovian decoherence. A simple example of qubit evolution is provided to illustrate our approach.  }

\section{Introduction}

A theory of open quantum systems provides a basic tool to analyze quantum systems which are not isolated  but  interact with  an external environment \cite{Breuer,Weiss,Rivas-Huelga,ALICKI}. Any realistic system is never perfectly isolated and hence this theory  plays a key role for modelling and controlling realistic quantum systems. It is, therefore, clear that open quantum systems are fundamental for potential applications in modern quantum technologies such as quantum communication, cryptography and computation \cite{QIT}. The standard approach \cite{Breuer} considers a total ``system + environment" Hamiltonian $\mathbf{H}$ and looks for the reduced evolution of the system density operator $\rho(t)$ defined by
\begin{equation}\label{RED}
 \rho_0 \ \longrightarrow\ \rho(t) = {\rm Tr}_E ( e^{-i \mathbf{H} t}\, \rho_0 \otimes \rho_E\, e^{i \mathbf{H} t}) ,
\end{equation}
where $\rho_E$ is an initial state of the environment and ${\rm Tr}_E$ denotes a partial trace over the environmental degrees of freedom. It is well known that the map $\rho_0 \longrightarrow \rho(t) = \Lambda_t[\rho_0]$ is completely positive and trace-preserving (CPTP) and satisfies $\Lambda_0 = \oper$ (identity map). It is usually called a (quantum) dynamical map. It was shown by Nakajima and Zwanzig \cite{NZ} (see also \cite{NZ-inni,Haake}) that $\rho(t)$ satisfies the following generalized master equation
\begin{equation}\label{NZ}
  \dot{\rho}(t) = \int_0^t \mathcal{K}_{t-\tau} \rho(\tau) d\tau ,
\end{equation}
in which quantum memory effects are taken into account through the introduction of the memory kernel $\mathcal{K}_t$. This means that the rate of change of the state $\rho(t)$ at time $t$ depends on its history (starting at $t$ = 0). The memory kernel is fully determined by the total Hamiltonian and the initial state of the environment. It should be stressed that in general its structure is highly nontrivial since the knowledge of the memory kernel derived from the microscopic model governed by the total Hamiltonian would be in principle equivalent to the knowledge of the full ``system + environment" evolution. The main problem of memory kernel master equation (\ref{NZ}) is the issue of complete positivity.  This problem was already posed by Barnett and Stenholm \cite{Stig}. An interesting approach of Shabani and Lidar \cite{Lidar} leads to so-called post-Markovian
master equation.  However, neither the phenomenological kernel of Barnett-Stenholm nor the Shabani-Lindar post-Markovian
kernel guarantees that the solution $\rho_0 \to \rho(t)$ defines a complete positive map (cf. also  \cite{Maniscalco}).
This problem was further extensively analyzed both from mathematical and physical point of view (see e.g. \cite{Budini,Wilke,AK,B-V,EPL,Wudarski,Bassano-PRL,PRA-2016,PRE-Breuer}). An interesting proposal leading to legitimate memory kernels is provided by so called collision models \cite{C0,C1,C2,C3,C4}. Actually, the non-local memory kernel master equation is well known for classical stochastic evolution \cite{Montroll,CLASS,Gil}, where the dynamical map is realized by a family of stochastic matrices. The quantum analog of semi-Markov evolution was proposed by Breuer and Vacchini \cite{B-V}, and then further analyzed in \cite{PRA-2016,Bassano-PRL} (see also recent papers \cite{Bassano2}). Very often one replaces Nakajima-Zwanzing equation (\ref{NZ}) by time-local Master equation

\begin{equation}\label{TL}
  \dot{\rho}(t) = \mathcal{L}_t\rho(t)
\end{equation}
with time-dependent local generator $\mathcal{L}_t$ \cite{Breuer,PRL-2010}. An interesting discussion on intricate relation between  (\ref{NZ}) and (\ref{TL}) can be found in \cite{Bassano1,Wegewijs}.

Why this problem is so hard? Note, that expressing the total Hamiltonian as

\begin{equation}\label{}
  \mathbf{H} = H_S \otimes \oper_E + \oper_E \otimes H_E + \mathbf{H}_{\rm int} ,
\end{equation}
with  the interaction Hamiltonian  given by

\begin{equation}\label{}
   \mathbf{H}_{\rm int} = \lambda \sum_\alpha A_\alpha \otimes B_\alpha ,
\end{equation}
where $\lambda$ denotes the coupling constant, one finds that the memory kernel $\mathcal{K}_t$ depends on all multi-time environmental correlation functions

\begin{equation}\label{}
C_{\alpha_1 \ldots \alpha_n}(t_1,\ldots,t_n) =   {\rm Tr}_E ( B_{\alpha_1}(t_1) \ldots B_{\alpha_n}(t_n) \rho_E) ,
\end{equation}
where $B_\alpha(t)= e^{i H_E t} B_\alpha e^{-i H_E t}$ are the time evolved environmental operators. Clearly, multi-time correlation functions contribute to non-Markovian memory effects of the quantum evolution governed by (\ref{NZ}).

For weak system-environment interaction ($\lambda \ll 1$) one usually applies well known Born approximation which allows to keep only 2-point function $C_{\alpha_1,\alpha_2}(t_1,t_2)$. If the initial environmental state $\rho_E$ is invariant w.r.t. free evolution $(\lambda=0)$, then all correlation function ar time homogeneous. In particular one has $C_{\alpha_1,\alpha_2}(t_1,t_2) = C_{\alpha_1,\alpha_2}(t_1-t_2)$.
Applying so called weak coupling approximation which assume not only Born approximation but also appropriate Markov approximation Davies \cite{Davies,DAVIES} derived Markovian Master equation

\begin{equation}\label{}
  \dot{\rho}(t) = \mathcal{L}\rho(t) ,
\end{equation}
where the time independent generator has the following well known structure

\begin{equation}\label{Davies}
  \mathcal{L} \rho = -i[H_S,\rho] +  \lambda^2 \mathbf{L} \rho ,
\end{equation}
with the dissipative part $\mathbf{L}$

\begin{widetext}
\begin{equation}\label{}
 \mathbf{L}\rho = -i [\Delta H_{\rm LS},\rho] +  \sum_{\alpha,\beta} \sum_{\omega}\gamma_{\alpha\beta}(\omega)  \left( A_\alpha(\omega) \rho A^\dagger_\beta(\omega) - \frac 12 \{ A^\dagger_\beta(\omega)A_\alpha(\omega),\rho \} \right) ,
\end{equation}
\end{widetext}
Moreover,  assuming the spectral representation for the system's Hamiltonian $H_S = \sum_k E_k P_k$

\begin{equation}\label{}
  A_\alpha(\omega) = \sum_{E_k-E_l=\omega} P_k A_\alpha P_l
\end{equation}
one finds  the Lamb shift correction

\begin{equation}\label{LS}
  \Delta H_{\rm LS} =  \sum_{\alpha,\beta} \sum_{\omega} s_{\alpha\beta}(\omega)  A^\dagger_\alpha(\omega)A_\beta(\omega) .
\end{equation}
Finally, $s_{\alpha\beta}(\omega)$ and $\gamma_{\alpha\beta}(\omega)$ are controlled by 2-point correlation function via

\begin{equation}\label{}
  \int_0^\infty e^{i \omega t} C_{\alpha\beta}(t) dt = \frac 12 \gamma_{\alpha\beta}(\omega) + i s_{\alpha\beta}(\omega) .
\end{equation}
The basic properties of 2-point correlation function imply that for each Bohr frequency $\omega = E_k - E_l$ the matrix $\gamma_{\alpha\beta}(\omega)$ is positive definite and hence due to the celebrated result of Gorini, Kossakowski, Sudarshan and Lindblad \cite{GKS,L}  Davies generator (\ref{Davies}) gives rise to CPTP evolution. Davies generators and weak cupling limit were further analyzed in \cite{Dumcke-1,Dumcke-2} (see also \cite{Roga} for more recent analysis of Davies generators for qutrits). Recently, Merkli \cite{Merkli} provided interesting improvement of Davies analysis.

Davies approach perfectly works whenever the system-environment interaction is weak and one has a clear separation of the
typical timescales associated with the system and the environment, i.e. one deals with fast environmental degrees of freedom and slow system's degrees of freedom. Appropriate averaging over fast degrees of freedom leads to Markovian Master equation for the slow degrees of freedom. This is a typical situation for quantum optical systems  (cf. \cite{Carmichael,Zoller}). However, for many important systems currently considered for modern quantum technologies the Born-Markov approximation is not justified and one has to consider more general approach beyond Markovian semigroup. In this paper we propose a hybrid approach. Quantum evolution governed by Davies generator describes three processes: unitary evolution (governed by the effective Hamiltonian $H_{\rm eff} = H_S + \Delta H_{\rm LS}$), decoherence and dissipation process. Our hybrid approach consists of replacing the Markovian dissipation process by more general one governed by memory kernel Master equation.

The paper is organized as follows: Section \ref{SEMI} introduces important class of classical non-Markovian processes - so called semi-Markov ones. Section \ref{Hybrid} provides the construction of the hybrid generator using semi-Markov classial process for diagonal elements of the density operator. This hybrid approach is illustrated in the qubit case in Section \ref{Qubit}. Finally, we conclude and provide an outlook in
Section \ref{Last}.

% $\rho_{ij}(t) = \lambda_{ij}(t) \rho_{ij}(0)$, with $\lambda_{ij}(t)|\leq 1$, and dissipation process $\rho_{ii}(t) = \sum_j T_{ij}(t) %\rho_{jj}(0)$, with $T_{ij}(t)$ being a stochastic matrix.

\section{Classical semi-Markov evolution}   \label{SEMI}

The classical analog of Markovian Master equation is a Pauli rate equation for the probability vector $\mathbf{p}(t)$

\begin{equation}\label{P1}
  \dot{\mathbf{p}}(t) = \mathrm{L} \mathbf{p}(t) ,
\end{equation}
where the classical generator $\mathbf{L}$ is represented by a real square matrix $\mathrm{L}_{kl}$ satisfying well known Kolmogorov conditions \cite{Kampen}

\begin{equation}\label{}
  \mathrm{L}_{kl} \geq 0 \ , \ \ \ \sum_k \mathrm{L}_{kl} = 1 ,
\end{equation}
and hence can be represented as follows

\begin{equation}\label{LW}
  \mathrm{L}_{kl} = W_{kl} - \delta_{kl} \sum_i W_{il} ,
\end{equation}
with $W_{kl}$ satisfying $W_{kl} \geq 0$ (note, that $W_{kk}$ does not affect $\mathrm{L}_{kl}$). In terms of $W_{kl}$ Pauli equation reads as follows

\begin{equation}\label{Pauli}
  \dot{p}_k(t) = \sum_{l} \Big( W_{kl} p_l(t) - W_{lk} p_k(t) \Big) .
\end{equation}
Going beyond Markovian scenario one replaces (\ref{P1}) by the corresponding memory kernel rate equation

\begin{equation}\label{}
  \dot{\mathbf{p}}(t) = \int_0^t \mathrm{K}_{t-\tau} \mathbf{p}(\tau) d\tau ,
\end{equation}
or equivalently

\begin{equation}\label{PW}
  \dot{p}_k(t) = \sum_{l} \int_0^t \Big( W_{kl}(t-\tau) p_l(\tau) - W_{lk}(t-\tau) p_k(\tau) \Big)d\tau ,
\end{equation}
with time dependent rates $W_{kl}(\tau)$. Now, contrary to the Markovian case, the condition for the rates $W_{kl}(\tau)$ which guarantee that the solution to (\ref{PW}) provides the stochastic matrix $\mathrm{T}(t)$

\begin{equation}\label{}
  \mathbf{p}(t) = \mathrm{T}(t) \mathbf{p}(0) ,
\end{equation}
are not known. There is, however, an interesting class of classical memory kernels which can be completely characterized and corresponds to so called semi-Markov classical processes \cite{Montroll,Gil,B-V,Esposito,CLASS}: one defines a semi-Markov matrix $q_{ij}(\tau) \geq 0$ for $\tau \geq 0$ such that $\int_0^t q_{ij}(\tau) d\tau$ denotes the probability of jump from state ``$j$" to state ``$i$" no later than  $\tau=t$ provided that at time $\tau=0$ the system stays at the state ``$j$". One introduces  waiting time distribution $f_j(\tau) =  \sum_i q_{ij}(\tau)$ and survival probability
\begin{equation}\label{}
  g_j(t) = 1 -  \int_0^t f_j(\tau)d\tau ,
\end{equation}
that is the probability that the system stays in the  state ``$j$"  up to $\tau=t$. Clearly $\sum_i \int_0^\infty q_{ij}(\tau) d\tau \leq 1$ and hence $g_j(t) \in [0,1]$.
%In particular $q_{ij}\tau) = \pi_{ij} f_j(\tau)$, where $\pi_{ij}$ is the transition probability (stochastic matrix).
Note, that the following diagonal matrix
\begin{equation}\label{n}
n_{ij}(\tau) = g_j(\tau) \delta_{ij} ,
\end{equation}
satisfies

\begin{equation}\label{}
  n_{ij}(t)\geq 0  , \ \ \ n_{ij}(0) = \delta_{ij} ,
\end{equation}
but it is not a stochastic matrix since $\sum_i n_{ij}(t)\leq 1$. Interestingly, one may `normalize' $n_{ij}(t)$ using the semi-Markov matrix $q_{ij}(t)$. One proves that the following matrix in the Laplace transform domain

\begin{equation}\label{}
  \widetilde{\mathrm{T}}(s) = \widetilde{n}(s)\, \frac{1}{\mathbb{I} - \widetilde{q}(s)} = \widetilde{n}(s)\, \sum_{n=0}^\infty \widetilde{q}^{n}(s) ,
\end{equation}
defines the stochastic matrix in the time domain

\begin{equation}\label{]}
   \mathrm{T}_{ij}(t) = n_{ij}(t) + (n \ast q)_{ij}(t) + (n\ast q \ast q)_{ij}(t) + \ldots ,
\end{equation}
provided that the above infinite series converges. In the above formula we use the following definitions of the Laplace transform and convolution

\begin{equation}\label{}
 \widetilde{f}(s) := \int_0^\infty e^{-s t} f(t) dt\ , \ \  (f \ast g)(t) := \int_0^t f(t-\tau) g(\tau) d\tau .
\end{equation}
It turns out that $\mathrm{T}(t)$ satisfies

\begin{equation}\label{T-semi}
  \dot{\mathrm{T}}_{ij}(t) = \int_0^t \sum_k \Big( W_{ik}(t-\tau) \mathrm{T}_{kj}(\tau) - W_{ki}(t-\tau)\mathrm{T}_{ij}(\tau) \Big) d\tau ,
\end{equation}
where the rates $W_{ij}(t)$ are defined by the following relation in the Laplace transform domain

\begin{equation}\label{w-q}
  \widetilde{W}_{ij}(s) = \frac{\widetilde{q}_{ij}(s)}{\widetilde{g}_j(s)} ,
\end{equation}
that is, $W_{ij}(t)$ is uniquely defined in terms of the semi-Markov matrix $q_{ij}(t)$.

\section{A hybrid generator}  \label{Hybrid}

A generic Davies generator has the following structure

\begin{equation}\label{}
  \mathcal{L} = \mathcal{L}_0 + \mathcal{L}^{\rm diss} + \mathcal{L}^{\rm dec} ,
\end{equation}
with $\mathcal{L}_0\rho = -i[H,\rho]$, where $H= \sum_k E_k  |k\rangle \langle k|$, and

\begin{equation}\label{}
   \mathcal{L}^{\rm diss} = \sum_{k\neq l} W_{kl} \Big( |k\rangle \langle l| \rho |l\rangle \langle k|  - \frac 12 \{ |l\rangle \langle l|,\rho\} \Big)
\end{equation}
and
\begin{equation}\label{}
  \mathcal{L}^{\rm dec}\rho = \sum_{k,l} D_{kl}  |k\rangle \langle k| \rho |l\rangle \langle l| - \frac 12 \sum_k D_{kk} \{  |k\rangle \langle k| ,\rho\} .
\end{equation}
Moreover,  $W_{kl} \geq 0$, and the matrix $D_{kl}$ is positive definite. Note, that

$$   [ \mathcal{L}_0, \mathcal{L}^{\rm diss}]=  [ \mathcal{L}_0, \mathcal{L}^{\rm dec}]=  [ \mathcal{L}^{\rm dec}, \mathcal{L}^{\rm diss}]=0 , $$
and hence the corresponding dynamical map reads

\begin{equation}\label{ttt}
  \Lambda_t = e^{t\mathcal{L}} = \mathbb{U}_t \circ \Phi^{\rm diss}_t \circ \Phi^{\rm dec}_t ,
\end{equation}
with

\begin{equation}\label{}
  \mathbb{U}_t \rho = e^{-iHt}\rho e^{iHt}\ , \ \ \Phi^{\rm diss}_t = e^{t \mathcal{L}^{\rm diss}}\ , \ \ \Phi^{\rm dec}_t = e^{t \mathcal{L}^{\rm dec}}\ .
\end{equation}
Each map $\mathbb{U}_t$, $ \Phi^{\rm diss}_t$, and  $\Phi^{\rm dec}_t $ is CPTP and so is their concatenation $\Lambda_t$. Moreover

\begin{equation}\label{}
  \Phi^{\rm diss}_t \rho = \sum_{k,l} \mathrm{T}_{kl}(t)  |k\rangle \langle l| \rho |l\rangle \langle k| + \sum_{k\neq l} \lambda_{kl}(t)  |k\rangle \langle k| \rho |l\rangle \langle l| ,
\end{equation}
where

\begin{equation}\label{lww}
  \lambda_{kl}(t) = \exp( - \frac 12 (w_k + w_l) t) \ , \ \ \ w_k := \sum_i W_{ik} ,
\end{equation}
and $\mathrm{T}_{kl}(t)$ is a stochastic matrix being a solution to the classical Pauli rate equation (\ref{Pauli}). Complete positivity of $ \Phi^{\rm diss}_t$ is guaranteed by the positivity of the following matrix

\begin{equation}\label{C}
  \mathbf{C}_{kk}(t) := \mathrm{T}_{kk}(t) \ , \ \ \ \mathbf{C}_{kl}(t) := \lambda_{kl}(t) , \ \ (k \neq l) .
\end{equation}
Similarly,

\begin{equation}\label{}
  \Phi^{\rm dec}_t \rho = \sum_k  |k\rangle \langle k| \rho |k\rangle \langle k| +  \sum_{k\neq l} \mu_{kl}(t)  |k\rangle \langle k| \rho |l\rangle \langle l| ,
\end{equation}
with

\begin{equation}\label{}
  \mu_{kl}(t) = e^{-t D_{kl}} , \ \ (k \neq l) .
\end{equation}
Hence

\begin{eqnarray}\label{LLL}
&& \mathbb{U}_t  \Phi^{\rm dec}_t \Phi^{\rm dec}_t \rho = \sum_{k,l} \mathrm{T}_{kl}(t)  |k\rangle \langle l| \rho |l\rangle \langle k| \nonumber \\ && +  \sum_{k\neq l} e^{-i(E_k-E_l)t} \lambda_{kl}(t)\mu_{kl}(t)  |k\rangle \langle k| \rho |l\rangle \langle l| .
\end{eqnarray}

Now, we propose the following generalization: we replace $\mathcal{L}^{\rm diss}$ by the corresponding memory kernel

\begin{equation}\label{}
   \mathcal{K}_t^{\rm diss}\rho = \sum_{k\neq l} W_{kl}(t) \Big( |k\rangle \langle l| \rho |l\rangle \langle k|  - \frac 12 \{ |l\rangle \langle l|,\rho\} \Big) ,
\end{equation}
where $W_{kl}(t)$ is constructed out of semi-Markov matrix $q_{kl}(t)$ via (\ref{w-q}). Note, that

$$   [ \mathcal{L}_0, \mathcal{K}_t^{\rm diss}]=  [ \mathcal{L}_0, \mathcal{L}^{\rm dec}]=  [ \mathcal{L}^{\rm dec}, \mathcal{K}_t^{\rm diss}]=0 , $$
and hence the corresponding dynamical map has a form (\ref{ttt}), where now the stochastic matrix $\mathrm{T}_{kl}(t)$ is a solution of (\ref{T-semi}) and
$\lambda_{kl}(t)$ solve the following equation

\begin{equation}\label{}
  \dot{\lambda}_{kl}(t) = - \frac 12 \int_0^t (w_k(t-\tau) + w_l(t-\tau)) \lambda_{kl}(\tau) d\tau ,
\end{equation}
with $\lambda_{kl}(0)=1$, which reduces to (\ref{lww}) in the Markovian case. The main difference between Markovian generator $ \mathcal{L}^{\rm diss}$ and semi-Markov memory kernel  $\mathcal{K}_t^{\rm diss} $ is that $ \mathcal{L}^{\rm diss}$ always generates a CPTP dynamical map $ \Lambda_t^{\rm diss}$, whereas it is not always true for the semi-Markov generator. The main result of this paper consists in the following observation

\begin{Theorem} Given a semi-Markov matrix $q_{kl}(t)$ one can always define a positive matrix $D_{kl}$ such that

\begin{equation}\label{}
  \dot{\rho}(t) = -i[H,\rho(t)] + \int_0^t \mathcal{K}^{\rm dis}_{t-\tau}\rho(\tau) d\tau +  \mathcal{L}^{\rm dec}\rho(t) ,
\end{equation}
gives rise to CPTP dynamical map $\Lambda_t$.
\end{Theorem}
Indeed, the role of pure decoherence generator is to make the matrix $\mathbf{C}$ defined in (\ref{C}) positive, that is, the corresponding dynamical map is completely positive iff

\begin{equation}\label{}
  \Phi^{\rm dec}_t[\mathbf{C}(t)]\geq 0 ,
\end{equation}
for all $t \geq 0$. Hence, if the matrix $\mathbf{C}(t)$ is already positive it remains positive under the action of decoherence map $\Phi^{\rm dec}_t$ for completely arbitrary decoherence matrix $D_{kl}$. However, if $\mathbf{C}(t)$ is not positive (for some $t$), then the role of  $\Phi^{\rm dec}_t$ is to suppress the off-diagonal elements of $\mathbf{C}(t)$ in order to make it positive.

\section{Stochastic Hamiltonian and semi-Markov dissipation}

Actually, the same effect of suppressing off-diagonal elements of $\mathbf{C}(t)$  may be achieved by a family of stochastic Hamiltonians $H(t) = \sum_k E_k(t) |k \rangle \langle k|$, with

\begin{equation}\label{}
  E_k(t) = E_k + \xi_k(t) ,
\end{equation}
where $\xi_k(t)$ represents a collection of white noise satisfying

\begin{equation}\label{}
  \langle \langle \xi_k(t) \rangle \rangle = 0 \ ,\ \ \  \langle \langle \xi_k(t)\xi_l(s) \rangle \rangle = \delta_{kl} \gamma_k \delta(t-s) .
\end{equation}
One finds for the dynamical map

\begin{equation}\label{}
\Lambda_t(\rho) =  \langle \langle\, U_\xi(t)  \Phi^{\rm diss}_t(\rho) U^\dagger_\xi(t)\, \rangle\rangle ,
\end{equation}
where

\begin{equation}\label{}
  U_\xi(t) = e^{-i \int_0^t H(\tau) d\tau } = \sum_k e^{-i E_k t} e^{-i \int_0^t \xi_k(\tau) d\tau} |k\rangle \langle k| .
\end{equation}
Hence, using the following property of the Gaussian noise

\begin{equation}\label{}
   \langle \langle\, e^{-i \int_0^t [\xi_k(\tau) - \xi_l(\tau)]d\tau } \, \rangle\rangle = e^{-\frac 12 \langle \langle\, [  \int_0^t [\xi_k(\tau) - \xi_l(\tau)]d\tau]^2 \rangle\rangle } \,  ,
\end{equation}
one finds that $\Lambda_t$ is represented by (\ref{LLL}) with

\begin{equation}\label{}
  \mu_{kl}(t) = e^{-(\gamma_k + \gamma_l)t} .
\end{equation}
Again, one can always find $\gamma_k$ such that $\Lambda_t$ is completely positive. One can in principle consider also time dependent rates $\gamma_k(t)$ which lead to $\mu_{kl}(t) = \exp( - \int_0^t (\gamma_k(\tau) + \gamma_l(\tau))d\tau$. Note, that due to the assumption that the noise is  Gaussian the decoherence is fully controlled by 2-point correlation only. One can also consider more general scenario, i.e. a noise which is not Gaussian, giving rise to decoherence controlled by higher order correlation functions.

\section{Example: qubit evolution}  \label{Qubit}

To illustrate our approach let us analyze a qubit evolution. One has $H= \frac{\omega}{2}\sigma$, with $\omega=E_1-E_0$,

\begin{eqnarray}\label{}
  \mathcal{K}_t^{\rm diss}\rho & =& k_+(t) \Big( \sigma_+ \rho\sigma_- - \frac 12 \{ \sigma_-\sigma_+,\rho\} \Big) \nonumber \\ &+&
  k_-(t) \Big( \sigma_- \rho\sigma_+ - \frac 12 \{ \sigma_+\sigma_-,\rho\} \Big)  ,
\end{eqnarray}
and finally the pure dephasing generator

\begin{equation}\label{L-z}
  \mathcal{L}^{\rm dec}\rho = \frac{\gamma_z}{2} (\sigma_z \rho\sigma_z - \rho) .
\end{equation}
Actually, authors of \cite{PRE-Breuer} analyzed memory kernel master equation

\begin{equation}\label{}
  \dot{\rho}(t) = \int_0^t  \mathcal{K}_{t-\tau}^{\rm diss}\rho(\tau)d\tau ,
\end{equation}
taking a semi-Markov matrix

\begin{equation}\label{}
  q(t) = \begin{pmatrix}
              0 & k_+(t) \\
              k_-(t) & 0
            \end{pmatrix} , \ \ k_\pm(t) = \kappa_\pm e^{-\gamma t} .
\end{equation}
It was found that indeed one finds legitimate stochastic matrix $\mathrm{T}_{ij}(t)$, however, in general the solution $\Phi^{\rm diss}_t$ violates complete positivity. Now, the role of purely dephasing generator   (\ref{L-z}) is to restore complete positivity for sufficiently large dephasing rate $\gamma_z$. Following \cite{PRE-Breuer} consider $k_\pm(t) = \kappa_\pm e^{- \gamma t}$, with the additional condition $\gamma^2 \geq \max\{ 4\kappa_- ,4 \kappa_-\}$. This choice provides legitimate classical semi-Markov evolution and one finds for the diagonal elements of the stochastic matrix

\begin{equation}\label{}
  \mathrm{T}_{00}(t) = \frac{\kappa_+}{\kappa_+ + \kappa_-} + \frac{\kappa_-}{\kappa_+ + \kappa_-} \, e^{- \gamma t /2} \Big[ \cosh(dt/2) + \frac\gamma d \sin(dt/2) \Big] ,
\end{equation}
and

\begin{equation}\label{}
  \mathrm{T}_{11}(t) = \frac{\kappa_-}{\kappa_+ + \kappa_-} + \frac{\kappa_+}{\kappa_+ + \kappa_-} \, e^{- \gamma t /2} \Big[ \cosh(dt/2) + \frac\gamma d \sinh(dt/2) \Big] ,
\end{equation}
with
\begin{equation}\label{}
  d = \sqrt{\gamma^2 - 4(\kappa_+ + \kappa_-)} .
\end{equation}
Now, the decoherence factor $\lambda_{01}(t)$ reads

\begin{equation}\label{}
  \lambda_{01}(t) = e^{- \gamma t /2} \Big[ \cosh(\overline{d}t/2) + \frac{\gamma}{\overline{d}} \sinh(\overline{d}t/2) \Big] ,
\end{equation}
with

\begin{equation}\label{}
  \overline{d} = \sqrt{\gamma^2 - 4(\kappa_+ + \kappa_-)} .
\end{equation}
Complete positivity is controlled by positivity of

\begin{equation}\label{}
  \mathbf{C}(t) = \begin{pmatrix}
              \mathrm{T}_{00}(t) & \lambda_{01}(t) \\
              \lambda_{01}(t) & \mathrm{T}_{11}(t)
            \end{pmatrix} .
\end{equation}
It was already observed in \cite{PRE-Breuer} that $\mathbf{C}(t)$ needs not be positive and hence the evolution in general is not CPTP. Figure 1 shows the plot of ${\rm det}\, \mathbf{C}(t)$. The blue curve shows that the evolution is evidently not CPTP. Now, we turn on decoherence generator $\mathcal{L}^{\rm dec}$ controlled by $\gamma_z$. Figure 1 shows that for small $\gamma_z$ the evolution still violates CP (yellow curve), however for $\gamma_z=1$ (green curve) the evolution is perfectly CP.

\begin{figure} \label{Fig-BB}
\includegraphics[width=9cm]{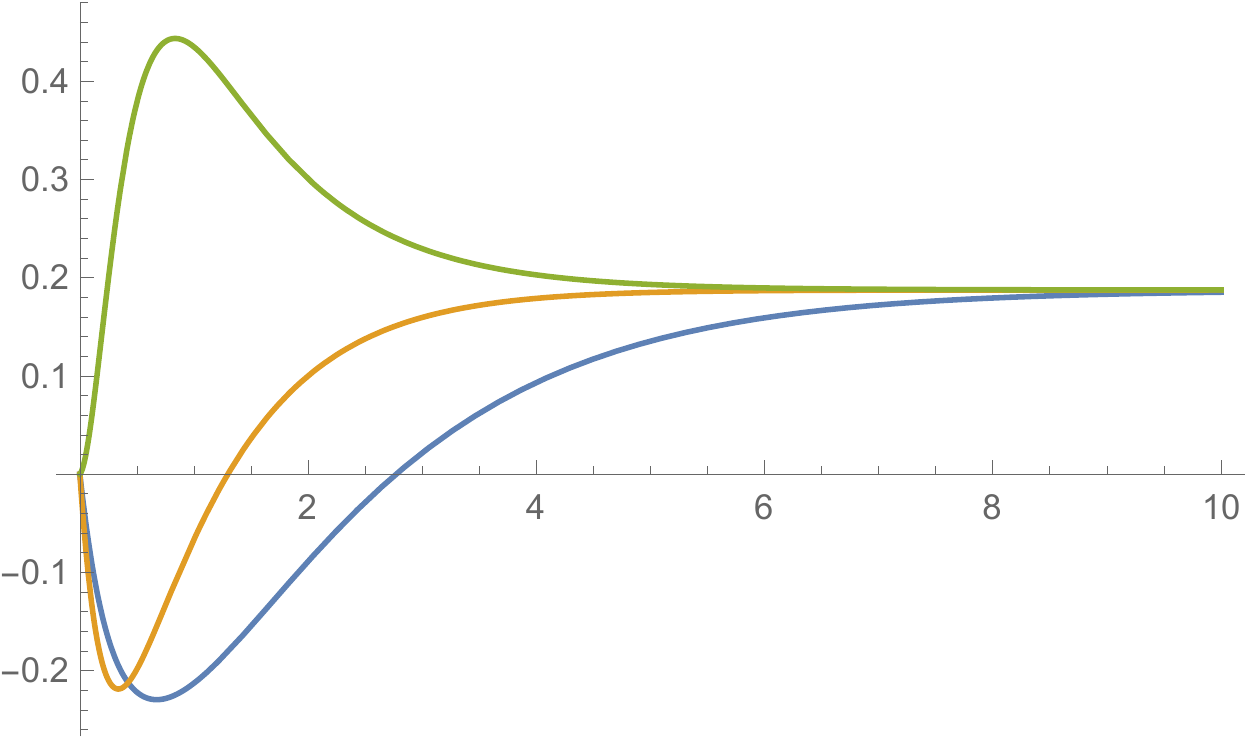}
\caption{(Color online) The plot of ${\rm det}\, \mathbf{C}(t)$. $\gamma=5$, $\kappa_1=1$, $\kappa_2=3$. Blue: $\gamma_z=0$, yellow: $\gamma_z=0.1$, green: $\gamma_z=1$. }
\end{figure}

\section{Conclusions}  \label{Last}

We provided a class of quantum evolution beyond Markovian semigroup. This class is governed by a hybrid Davies like generator: dissipation is controlled by a memory kernel and decoherence by standard GKLS generator. These two processes commute and both of them commute with the unitary evolution controlled by the Hamiltonian $H$. The corresponding memory kernel gives rise to semi-Markov evolution of the diagonal elemens of the density matrix. However, the corresponding map $\Phi^{\rm diss}_t$ needs not be completely positive (actually, in general it is not even positive). The role of decoherence generator $\mathcal{L}^{\rm dec}$ is to restore complete positivity. Hence, to pose the dynamical problem one needs that two processes generated by $\mathcal{K}_t^{\rm diss}$ and $\mathcal{L}^{\rm dec}$ are properly adjust. They are no longer independent as one has for a semigroup evolution. However, given $\mathcal{K}_t^{\rm diss}$  one can always find  $\mathcal{L}^{\rm dec}$ which properly suppresses all terms violating complete positivity. This simple scheme is illustrated for a qubit evolution.  Note, that presented approach may be easily generalized for time-dependent decoherence matrix $D_{kl}(t)$ and time-dependent Hamiltonian $H(t) = \sum_k E_k(t) |k\rangle \langle k|$. In particular modeling a decoherence process such that

\begin{equation}\label{}
  \mathrm{T}_{kk}(t) \mathrm{T}_{ll}(t) = |\lambda_{kl}(t) \exp( \int_0^t D_{kl}(\tau) d\tau ) |^2
\end{equation}
one realizes an evolution which maximally preserves the coherence compatible with the dissipation process encoded in $\mathrm{T}_{kl}(t)$ (cf. \cite{Karol-NJP}). It would be interesting to further analyze this hybrid approach in connection to quantum non-Markovianity (cf. recent reviews \cite{Rev1,Rev2,Rev3}). Clearly, a single qubit example serves as a simple illustration of the general method. It would be interesting to analyze a hybrid approach to many body quantum systems. We postpone this for further research.

\section*{Acknowledgements}

I thank K. \.Zyczkowski and S. Denisov for interesting discussions on the structure of GKLS generators. This paper was partially supported by the National Science Center project 2018/30/A/ST2/00837.

\section*{Data availability} The data that supports the findings of this study are available within the article.


\begin{thebibliography}{1} \bibliographystyle{plain}



\bibitem{Breuer} H.-P. Breuer and F. Petruccione,
{\em The Theory of Open Quantum Systems} (Oxford Univ. Press,
Oxford, 2007).

\bibitem{Weiss} U. Weiss, {\it Quantum Dissipative Systems}, (World
Scientific, Singapore, 2000).

\bibitem{Rivas-Huelga} \'A. Rivas and S. F. Huelga, {\it Open Quantum Systems. An Introduction}, Springer Briefs in Physics,
Springer 2011.


\bibitem{ALICKI} R. Alicki and K. Lendi, {\it Quantum Dynamical
Semigroups and Applications} (Springer, Berlin, 1987).

%\bibitem{Alicki} R. Alicki and K. Lendi, {\it Quantum Dynamical
%Semigroups and Applications} (Springer, Berlin, 1987).


\bibitem{QIT} M. A. Nielsen and I. L. Chuang, {\it Quantum
Computation and Quantum Information} (CUP,
Cambridge, 2000).

\bibitem{NZ} S. Nakajima, Prog. Theor. Phys. {\bf 20}, 948 (1958); R.
Zwanzig, J. Chem. Phys. {\bf 33}, 1338 (1960).



\bibitem{NZ-inni} S. Chaturvedi and J. Shibata, Z. Phys. B {\bf 35}, 297 (1979); N. H. F. Shibata and Y. Takahashi,
J. Stat. Phys. {\bf 17}, 171 (1977).


\bibitem{Haake} F. Haake, {\em Statistical Treatment of Open Systems by Generalized Master Equations},
(Springer Tracts in Modern Physics) vol. 66 (Berlin: Springer 1973).


\bibitem{Stig} S. M. Barnett and S. Stenholm, Phys. Rev. A {\bf 64}, 033808 (2001).


\bibitem{Lidar}  A. Shabani and D.A. Lidar, Phys. Rev. A {\bf 71}, 020101(R) (2005).


\bibitem{Maniscalco} S. Maniscalco, Phys. Rev. A {\bf 72}, 024103 (2005); S. Maniscalco and F. Petruccione, Phys.
Rev. A {\bf 73}, 012111 (2006).

%\bibitem{Campbell} S. Campbell, et al Phys. Rev. A {\bf 85}, 032120 (2012).


\bibitem{Budini} A. A. Budini, Phys. Rev. A {\bf 69}, 042107 (2004).

\bibitem{Wilke}   J. Wilkie,  Phys. Rev. E {\bf 62},  8808 (2000); J. Wilkie and Y.M.
Wong, J. Phys. A: Math. Theor. {\bf 42}, 015006 (2009);

\bibitem{AK} A. Kossakowski and R. Rebolledo, Open Syst. Inf. Dyn. {\bf 14}, 265 (2007); \textit{ibid.}
{\bf 16}, 259 (2009).



\bibitem{B-V} H.-P. Breuer and B. Vacchini, Phys.
Rev. Lett. {\bf 101} (2008) 140402.


\bibitem{PRE-Breuer} H.-P. Breuer and B. Vacchini, Phys. Rev. E {\bf 79},  041147 (2009).

\bibitem{EPL} D. Chru\'sci\'nski and A. Kossakowski, EPL {\bf 97}, 20005 (2012).

\bibitem{Wudarski} F. A. Wudarski, P. Nale\.zyty, G. Sarbicki, and D. Chru\'sci\'nski,
Phys. Rev. A {\bf 91}, 042105 (2015).

\bibitem{PRA-2016} D. Chru\'sci\'nski and A. Kossakowski,  Phys. Rev. A {\bf 94}, 020103(R) (2016).


\bibitem{Bassano-PRL} B. Vacchini, Phys. Rev. Lett. {\bf 117}, 230401 (2016).


%=========inne===============




\bibitem{C0} V. Scarani, et al. Phys. Rev. Lett. 88, 097905 (2002); M. Ziman, P. \v{S}telmachovi\v{c}, V. Bu\v{z}ek, M. Hillery, V. Scarani, and N. Gisin, Phys. Rev. A65, 042105,
(2002);  M. Ziman and V. Bu\v{z}ek, Phys. Rev. A {\bf 72}, 022110, (2005).


\bibitem{C1}  V. Giovannetti and G. M. Palma, Phys. Rev. Lett. {\bf 108}, 040401 (2012).

\bibitem{C2} F. Ciccarello, G. M. Palma, V. Giovannetti, Phys. Rev. A {\bf 87}, 040103(R) (2013).

\bibitem{C3} B. Vacchini, Phys. Rev. A {\bf 87}, 030101(R) (2013);  B. Vacchini,  Int. J. Quantum Inform.  {\bf 12}, 1461011 (2014).

\bibitem{C4} S. Lorenzo, F. Ciccarello, and G. M. Palma, Phys. Rev. A {\bf 93}, 052111 (2016).



\bibitem{Montroll} E. W. Montroll and G. H. Weiss, J. Math. Phys. {\bf 6}, 167 (1965); V. M. Kenre, E. W Montroll, M. F Shlesinger,  J. Stat. Phys. {\bf 9}, 45 (1973).

\bibitem{Gil} D. T. Gillespie, Phys. Lett. A, {\bf 64}, 22 (1977).

\bibitem{Esposito} M. Esposito and K. Lindenberg, Phys. Rev. E {\bf 77}, 051119 (2008).


\bibitem{CLASS} B. Vacchini, A. Smirne, E.-M. Laine, J. Piilo, and H.-P. Breuer,  New J. Phys. {\bf 13}, 093004 (2011).



\bibitem{Bassano1} N. Megier, A. Smirne, and B. Vacchini,  New J. Phys. {\bf 22}, 083011 (2020).



\bibitem{Bassano2} N. Megier, A. Smirne, and B. Vacchini, Entropy {\bf 22}, 796 (2020).

\bibitem{PRL-2010} D. Chru\'sci\'nski and A. Kossakowski, Phys. Rev. Lett. {\bf 104}, 070406 (2010).

\bibitem{Wegewijs}    K. Nestmann, V. Bruch, and M. R. Wegewijs, {\em How quantum evolution with memory is generated
in a time-local way},  arXiv:2002.07232.


%==============SEmi




\bibitem{Davies} E.B. Davies, Comm. Math. Phys. {\bf 39}, 91 (1974).

\bibitem{DAVIES} E. B. Davies, {\em Quantum Theory of Open Systems},  (Academic Press, London, 1976) .

\bibitem{GKS} V. Gorini, A. Kossakowski, and E. C. G. Sudarshan, J. Math. Phys.
{\bf 17}, 821 (1976).

\bibitem{L}  G. Lindblad, Comm. Math. Phys. {\bf 48}, 119
(1976).


\bibitem{Dumcke-1} R. D\"umcke and H. Spohn, Z. f. Phys. B  {\bf 34}, 419 (1979).

\bibitem{Dumcke-2} R. D\"umcke, J. Math. Phys. {\bf 24}, 311 (1983).


\bibitem{Roga} W. Roga, M. Fannes, and K. \.Zyczkowski, Rep. Math. Phys. {\bf 66}, 311 (2010)

\bibitem{Merkli} M. Merkli, Ann.  Phys. {\bf 412}, 167996 (2020).

%====== GKSL =========


\bibitem{Carmichael} H. J. Carmichael, {\em Statistical Methods in Quantum Optics 1: Master Equations and Fokker-Planck Equations}, Springer (2008).


\bibitem{Zoller} C. Gardiner and P. Zoller, {\em Quantum Noise: A Handbook of Markovian and Non-Markovian Quantum Stochastic Methods with with Applications to Quantum Optics}, (Springer Series in Synergetics) 3rd Edition (2004).


\bibitem{Kampen} N. G. van Kampen, Stochastic Processes in Physics and
Chemistry, North Holland, Amsterdam 2007.
%===============MARKOV=========

%\bibitem{BLP} H.-P. Breuer, E.-M. Laine, J. Piilo, Phys. Rev. Lett. {\bf 103},
%210401 (2009).




%==========Memory kernels======================









%=====================RECENT================

\bibitem{Karol-NJP} K. Korzekwa, S. Czach\'orski, Z. Pucha{\l}a, and K. \.Zyczkowski, New J. Phys. {\bf 20}, 043028
(2018).


\bibitem{Rev1} \'A. Rivas, S. F. Huelga, and M. B. Plenio,  Rep. Prog. Phys. {\bf 77}, 094001 (2014).

\bibitem{Rev2}  H.-P. Breuer, E.-M. Laine, J. Piilo, and B. Vacchini,  Rev. Mod.
Phys. {\bf 88}, 021002 (2016).

\bibitem{Rev3}  I. de Vega and D. Alonso, Rev. Mod. Phys. {\bf 89}, 015001
(2017).


%=========================

%\bibitem{Wod} S. Daffer, et. al.
%K. W\'odkiewicz, J.D. Cresser, and J.K. McIver,
%Phys. Rev. A {\bf 70}, 010304 (2004).



%==============COL=========







%\bibitem{Bassano} B. Vacchini,  Phys. Rev. A {\bf 87}, 030101(R) (2013).

%\bibitem{Laura}L. Mazzola, E.-M. Laine, H.-P. Breuer, S. Maniscalco, and J. Piilo, Phys. Rev. A {\bf81}, 062120 (2010).



%=====Collisonal============

%

%\bibitem{COL} P. Filipowicz, et. al. Phys. Rev. A {\bf 34}, 3077
%(1986); J. D. Cresser, Phys. Rev. A {\bf 46}, 5913 (1992); J. Cresser and S. Pickles, Quantum Semicl.
%Opt. {\bf 8}, 73 (1996); U. Herzog and J. Bergou, Phys. Rev. A {\bf 54}, 5334 (1996).





%=================================











\end{thebibliography}
\end{document}